%% file: main.tex
\documentclass{article}
\usepackage{spconf}
\usepackage{amsmath}
\usepackage{xcolor}
\usepackage{booktabs}
\usepackage{multirow}
\usepackage{csquotes}
\usepackage{graphicx}

\title{Cross-speaker style transfer for text-to-speech\\using data augmentation}

\name{Manuel Sam Ribeiro, Julian Roth, Giulia Comini, Goeric Huybrechts, Adam Gabryś, Jaime Lorenzo-Trueba}
\address{
    Amazon Alexa, TTS Research\\
    \fontsize{9}{9}\selectfont\ttfamily\upshape
    \{manuerib, huybrech, truebaj\}@amazon.com
    \vspace{-10pt}
    }

\begin{document}
\ninept
\maketitle
\begin{abstract}
We address the problem of cross-speaker style transfer for text-to-speech (TTS) using data augmentation via voice conversion.
We assume to have a corpus of neutral non-expressive data from a target speaker and supporting conversational expressive data from different speakers.
Our goal is to build a TTS system that is expressive, while retaining the target speaker’s identity. 
The proposed approach relies on voice conversion to first generate high-quality data from the set of supporting expressive speakers.
The voice converted data is then pooled with natural data from the target speaker and used to train a single-speaker multi-style TTS system.
We provide evidence that this approach is efficient, flexible, and scalable.
The method is evaluated using one or more supporting speakers, as well as a variable amount of supporting data.
We further provide evidence that this approach allows some controllability of speaking style, when using multiple supporting speakers.
We conclude by scaling our proposed technology to a set of 14 speakers across 7 languages.
Results indicate that our technology consistently improves synthetic samples in terms of style similarity, while retaining the target speaker's identity.
\end{abstract}

\input{body.tex}

\bibliographystyle{IEEEbib}
\bibliography{references}

\end{document}

%% file: body.tex
\section{Introduction}

Text-to-speech (TTS) technology is consistently reducing the perceived gap between synthetic and natural speech.
This is being achieved with the development of novel acoustic modeling \cite{shen2018natural, tan2021survey} and waveform generation techniques \cite{tamamori2017speaker, lorenzo2019towards}.
Leveraging these methods, researchers have been focusing on the flexibility and controllability of TTS systems, especially for expressive data \cite{lee2021styler, shechtman2021synthesis, valle2020mellotron}.
The development of flexible systems requires the ability to control speaking style and speaker identity, among other speech attributes.
Speaking style denotes the global attributes that describe the emotion, affect, and/or generic attitude conveyed through speech by a speaker in a particular domain.
For example, we may define speaking styles such as read, newscaster, conversational, or emotional speech.
The ability to control these attributes is essential for TTS voices that are flexible and adaptable to multiple scenarios and domains.
Traditionally, extending TTS voices to new domains where speaking style is relevant involved additional data collection from the target voice.
This method, however, does not scale well, as it is not always feasible to record more data for a specific voice talent.
An alternative approach is to transfer speaking style from other speakers for which recorded data is already available.

Cross-speaker style transfer involves the generation of speech samples that are perceived to have the identity of a target speaker and the speaking style of a supporting speaker.
To control speaking style, recent work proposed the inclusion of reference encoders \cite{skerry2018towards, wang2018style}.
Together with textual input, the TTS model inputs a reference speech representation that is used to condition the generated speaking style.
Frequently used methods for the reference encoder are based on Global Style Tokens (GST) \cite{wang2018style} or Variational Auto-Encoders (VAEs) \cite{kingma2013auto, zhang2019learning}.
Although these methods could be used for cross-speaker style transfer, they have some shortcomings.
They may be dependent on similar text, where the reference waveform is similar to the textual input to be synthesized \cite{skerry2018towards, sorin2020principal}.
Or they might not model the target speech attributes, especially in disjoint corpora, without data from the target speaker in the target speaking style \cite{whitehill2019multi}.

For these reasons, most systems for cross-speaker style transfer aim to explicitly disentangle speaker identity and speaking style from other speech attributes.
This is primarily accomplished with multi-speaker TTS models \cite{valle2020mellotron}.
For speech generation, the model is conditioned on the target speaker identity and the desired speaking style.
Recent studies proposed the usage of multiple reference encoders, with each encoder modeling a specific speech attribute.
These systems can be trained using intercross or adversarial training \cite{bian2019multi, whitehill2019multi}.
Models are typically optimized on a variety of loss functions, such as cycle consistency loss \cite{xue2021cycle}, adversarial consistency loss \cite{whitehill2019multi, an2021improving}, N-pair loss \cite{kulkarni2021improving}
or other loss functions defined over latent representations of speaker identities and/or speaking styles \cite{joo2020effective, an2021improving}.
Instead of multiple encoders, hierarchical architectures were also proposed \cite{an2019learning}.
Alternatively, control of speaking style may be left to models of \emph{f0} and duration, learned separately or implicitly with the TTS model \cite{valle2020mellotron, shechtman2021synthesis, pan2021cross}.

A related area of research focuses on the development of TTS voices for low-resource scenarios.
Such methods aim to synthesize speech given a limited amount of training data from the target speaker, domain, or language.
Recent studies proposed to augment small corpora with synthetic data, either via voice conversion \cite{huybrechts2021low, shah2021non} or a large text-to-speech system \cite{hwang2021tts}.
These studies provided evidence that high-quality synthetic data can be used efficiently to complement natural data.

In this paper, we propose to address the problem of \emph{cross-speaker style transfer for text-to-speech using data augmentation}.
We assume to have a corpus of neutral non-expressive data from a target speaker and some supporting conversational expressive data from other speakers.
Our goal is to build a TTS system that is expressive, while retaining the target speaker's identity.
Our method uses voice conversion to generate high-quality data from a set of supporting expressive speakers.
The synthetic expressive data is pooled with the natural non-expressive data to train a single-speaker multi-style TTS system.
We show that our proposed method is flexible and capable of transferring speaking style across speakers.
Additionally, we show the effectiveness of our proposed method when
1) using data from one or more supporting speakers;
2) using as little as 1 hour of supporting data;
3) controlling the speaking style via the TTS reference encoder's latent space; and
4) applied to a variety of languages and speakers.

\section{Model Architecture}
\label{sec:methodology}

\subsection{Voice Conversion}

\begin{figure}[t]
  \centering
  \includegraphics[width=1.0\linewidth]{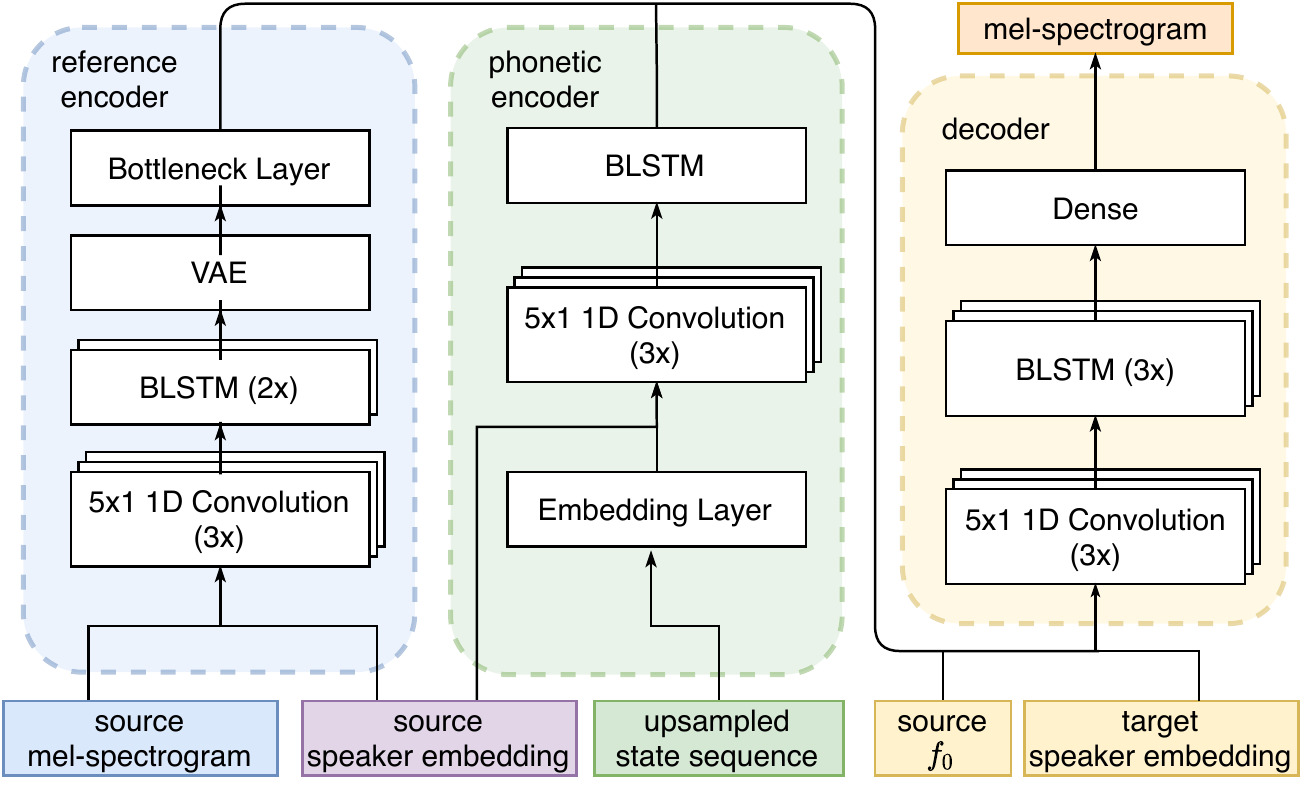}
  \caption{Voice Conversion model architecture.}
  \label{fig:vc_model}
\end{figure}

Our voice conversion approach (Figure \ref{fig:vc_model}) is based on CopyCat \cite{karlapati2020copycat},
extended to be conditioned on the source utterance's \emph{log-f0} \cite{qian2020f0}.
The purpose of this architecture is to preserve linguistic content and prosodic attributes, while modifying only speaker identity.
The reference encoder inputs a source speaker's mel-spectrogram and it includes a bottleneck layer that downsamples and upsamples the latent representations along the time dimension \cite{qian2019autovc}.
The phonetic encoder's architecture follows the Tacotron 2 text encoder \cite{shen2018natural}.
It inputs a categorical representation of Hidden Markov Model (HMM) states.
The HMM state sequence is found by force-aligning the training data at the phone level to the corresponding Mel-Frequency Cepstral Coefficient sequence.
We use 3-state left-to-right HMMs for each individual phone.
To match the time resolution of the mel-spectrogram, each time-aligned HMM state is upsampled to the frame level.
Both reference and phonetic encoders further input a speaker embedding defined at the utterance level, which is broadcasted to the number of frames in each utterance \cite{huybrechts2021low}.
Speaker embeddings are learned on a large multi-speaker multi-lingual corpus and optimized on a Generalized End-to-End Loss \cite{wan2018generalized}.
The encoded reference and phonetic sequences are concatenated with the utterance's \emph{log-f0} and speaker embedding.
The model is optimized on data from the target and supporting speakers for 100k steps using a batch size of 32 utterances.
We use a KL-divergence loss for the VAE component and an L1 loss on the source and reconstructed mel-spectrograms.
We further fine-tune the model for 25k steps only on training data from the target speaker.

For voice conversion, the reference and phonetic encoders input data from the source speaker, while the decoder is conditioned on information from the target speaker.
The target speaker embedding is the centroid of all embeddings from that speaker's training data.
The observed source utterance's \emph{f0} is mean-normalized to the target speaker's mean \emph{f0}.
The last layer of the decoder transforms the input frame-by-frame, using contextual information provided by the recurrent layers.
Note that there is no attention or alignment required in this architecture, as we preserve the source utterance's duration.
With the addition of the source \emph{f0} signal, the reconstructed mel-spectrogram preserves the prosodic properties describing the source style.
Any other relevant information not accounted for by duration, \emph{f0}, or speaker identity, is carried-over by the reference encoder.

\subsection{Text-to-Speech}

\begin{figure}[t]
  \centering
  \includegraphics[width=1.0\linewidth]{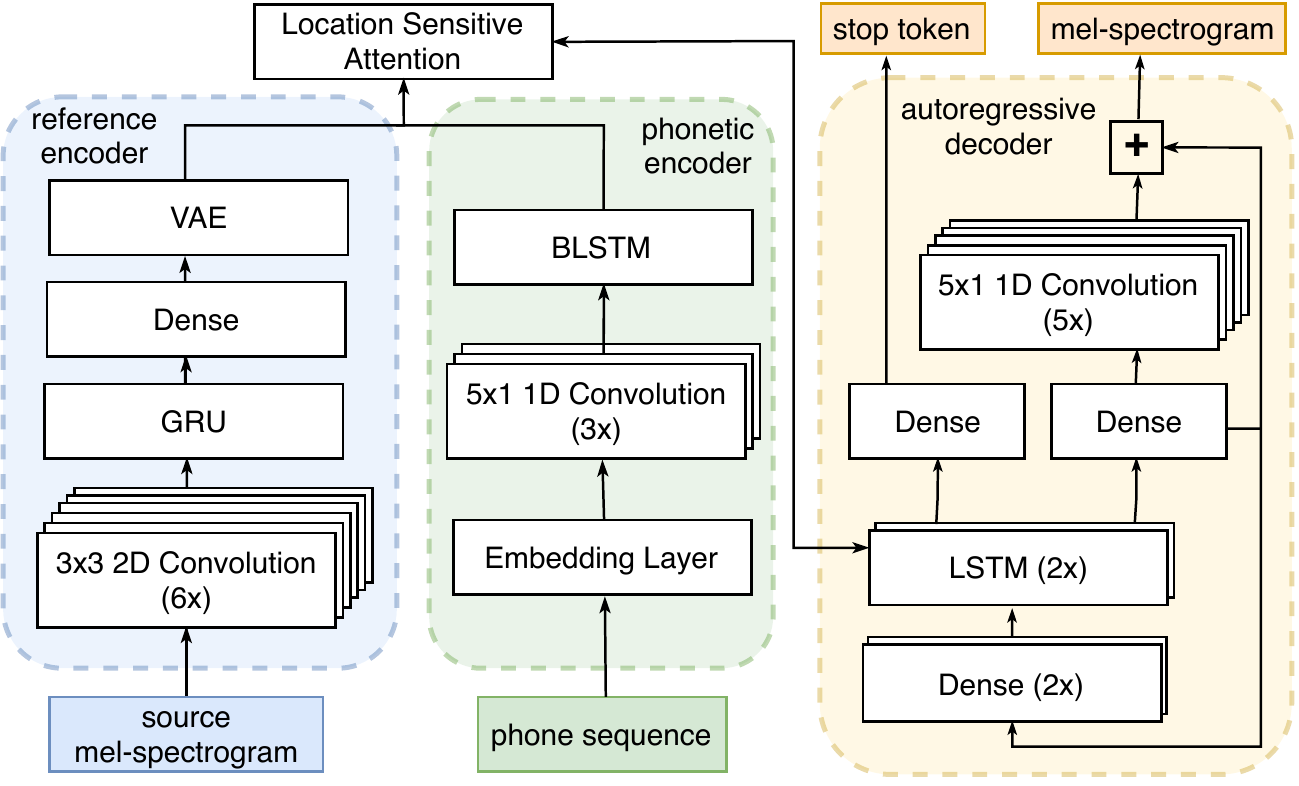}
  \caption{Text-to-Speech model architecture.}
  \label{fig:tts_model}
\end{figure}

Our TTS model (Figure \ref{fig:tts_model}) is a sequence-to-sequence encoder-decoder architecture based on Tacotron 2 \cite{shen2018natural}, using a single-head location-sensitive attention mechanism \cite{bahdanau2014neural}.
The phonetic encoder and decoder follow the architecture proposed by Shen et al \cite{shen2018natural}.
The reference encoder follows the architecture proposed by Skerry-Ryan et al \cite{skerry2018towards} with the addition of a Variational Auto-Encoding layer \cite{kingma2013auto}.
TTS systems are trained for 400k steps using a batch size of 32 utterances and optimized on an L1 loss for the generated mel-spectrograms, a KL-divergence loss for the Variational Autoencoder, and a cross-entropy loss for the stop-token.
The phonetic sequence is extracted from language-specific front-ends and corresponds to phone identities, word boundary tokens, and stress markers.

For style transfer TTS systems, we pool the synthetic expressive mel-spectrograms generated by the voice conversion model with the natural non-expressive mel-spectrograms from the target speaker.
Because speaker identity has been modified by voice conversion, there is no need to train multi-speaker models or use multiple reference encoders.
To generate speech samples, we condition the decoder on a VAE z-vector computed over the training data converted from a single supporting speaker, corresponding to a unique speaking style.
We finally convert the mel-spectrograms generated by the TTS models to time-domain waveforms using a Parallel Wavenet universal neural vocoder \cite{jiao2021universal}.

\section{Experiments}

Our goal is to build conversational text-to-speech systems for voices that only have speech recordings read in a \emph{neutral speaking style}.
This speaking style is characterized by flat, monotonous, and unexpressive read speech.
As supporting data, we use recordings from different speakers in a \emph{conversational speaking style}, which aims to capture a natural, friendly, and expressive speaking style.
Throughout our experiments, we use the terms Target, Supporting, and Source speaker.
\enquote{Target speaker} denotes the speaker for which neutral data is available, and the speaker identity we wish to preserve in the synthetic samples.
\enquote{Supporting speakers} indicates the set of speakers from which we draw supporting conversational data used to augment the TTS models.
\enquote{Source speaker} denotes the supporting speaker used to condition the TTS model when generating speech samples.
We compute a VAE z-vector centroid over the set of voice-converted training samples from the Source speaker.

\subsection{Supporting speakers}
\label{subsec:supporting_speakers}

\begin{table}[t]
\centering
\resizebox{1.0\columnwidth}{!}{%
\begin{tabular}{lccc}
\toprule
\textbf{System}     & \textbf{Naturalness}  & \textbf{Speaker Sim.} & \textbf{Style Sim.}           \\ \midrule
Recordings          &  61.51 $\pm$ 1.53     &    -                  & -                   \\
Neutral             &  54.76 $\pm$ 1.39     &    71.93 $\pm$ 1.54   & 38.12 $\pm$ 1.65    \\
Augmented (1 spk)   &  59.07 $\pm$ 1.37     &    64.56 $\pm$ 1.57   & 58.60 $\pm$ 1.58    \\
Augmented (4 spks)  &  58.98 $\pm$ 1.39     &    65.03 $\pm$ 1.55   & 60.05 $\pm$ 1.55    \\
Augmented (8 spks)  &  59.51 $\pm$ 1.36     &    64.54 $\pm$ 1.57   & 59.17 $\pm$ 1.57    \\
Source Speaker      &  -                    &    27.85 $\pm$ 1.61   & 72.81 $\pm$ 1.59    \\ \bottomrule
\end{tabular}%
}
\caption{MUSHRA evaluation in terms of Naturalness, Speaker Similarity, and Speaking Style Similarity. Results indicate mean score with 95\% confidence interval. \enquote{Augmented} systems denote target speaker TTS systems augmented with conversational data from a number of supporting speakers. \enquote{Source speaker} indicates a conversational TTS system from the source supporting speaker.}
\label{table:exp_supporting_speakers}
\end{table}

To investigate our proposed cross-speaker style transfer approach, we use internal corpora of Brazilian Portuguese data, choosing the target speaker to be a female speaker and supporting speakers to be gender-balanced.
\emph{Neutral} denotes a system trained on 10 hours of neutral data from the target speaker.
\emph{Augmented} systems indicate models trained using data augmentation via voice conversion.
For this experiment, we convert a total of 8 hours of conversational data to the identity of the target speaker.
We keep the amount of supporting data fixed and we vary the number of supporting speakers.
The system with 1 supporting speaker uses 8 hours of data from a single speaker, while the systems using 4 and 8 supporting speakers use 2 hours and 1 hour of data from each supporting speaker, respectively.
The source speaker is kept constant across all augmented systems.
For comparison, \emph{Source} indicates a TTS system trained on the 8 hours of conversational data from the source speaker.
We evaluate our systems with a set a MUSHRA-like (MUltiple Stimuli with Hidden Reference and Anchor) \cite{series2014method} listening tests.
Naturalness is evaluated omitting the reference and including a recording sample with the competing systems.
We further evaluate systems in terms of Speaker Similarity and Style Similarity.
The speaker similarity evaluation uses a reference sample drawn from the training data of the target speaker, while the style similarity evaluation uses a reference sample drawn from the training data of the source speaker.
For these tests, listeners were asked to rate the samples based only on the similarity of the speaker identity or the speaking style, ignoring all other attributes.
Each listening test included 150 utterances generated by each of the competing systems.
Utterances were rated by 100 native speakers using a crowdsourcing platform.
Each listener rated no more than 15 MUSHRA screens.

\textbf{Results}  in Table \ref{table:exp_supporting_speakers} show MUSHRA mean scores with a 95\% confidence interval.
We observe that the proposed Augmented systems improve naturalness over the Neutral system.
In terms of speaker similarity, the Augmented systems score below the Neutral system, although this is somewhat expected due to the differences in speaking style between competing and reference samples.
On average, however, the Augmented systems bridge the gap between source and target speakers by 83.6\%.
For style similarity, the Augmented systems outperform the neutral system, suggesting that listeners are able to discern between the speaking style generated by the competing systems.
For each subjective evaluation, we perform paired two-sided t-tests on the MUSHRA scores with a Holm-Bonferroni correction for multiple comparisons.
Across the three evaluated dimensions, we observe no statistically significant difference between Augmented systems at the level of $p<.01$.

\subsection{Controllability}
\label{subsec:controllability}

\begin{figure}[t]
  \centering
  \includegraphics[width=0.96\linewidth]{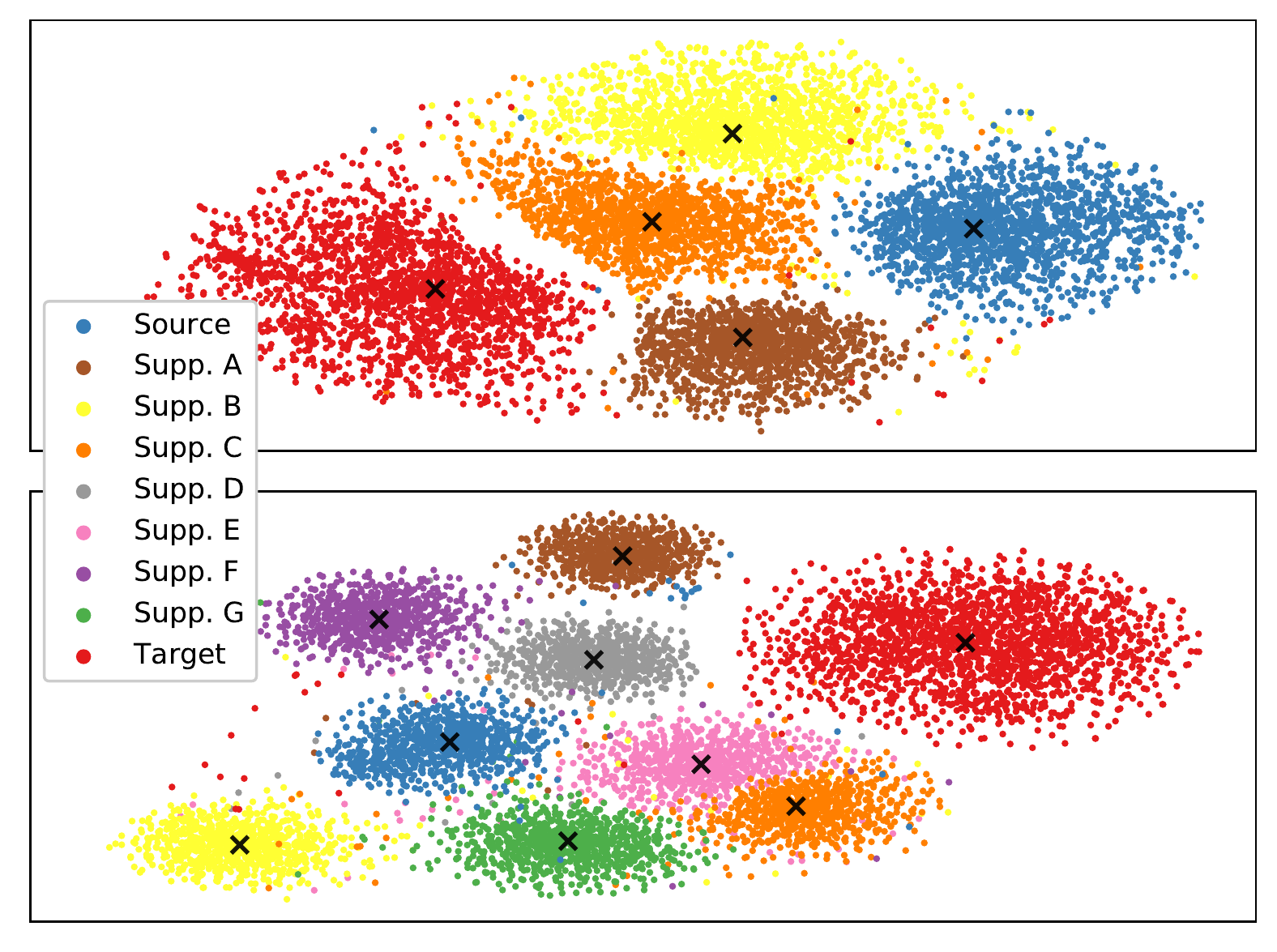}
  \caption{VAE space for systems augmented with voice-converted data from 4 (top) and 8 supporting speakers (bottom), visualized with t-SNE. The centroid of the training data converted from each speaker is marked with \textbf{X}.}
  \label{fig:vae_space}
\end{figure}

\begin{table}[t]
\centering
\resizebox{0.97\columnwidth}{!}{%
\begin{tabular}{@{}cc|c|c|c|c@{}}
\toprule
    & \multicolumn{5}{c}{\textbf{Reference}}     \\ \cmidrule(l){2-6}
 \parbox{1.5cm}{\centering \textbf{Centroid}}  &  \textbf{Target}      & \textbf{Source}       &   \textbf{Supp. A}     & \textbf{Supp. B}     & \textbf{Supp. C}     \\ \midrule
\textbf{Target}        &  \textbf{48.28\%}  &  6.55\%   &   14.33\%  &  5.52\%  & 10.74\%  \\
\textbf{Source}        &  13.45\%           &  \textbf{32.07\%}  &   28.33\%  &  22.07\% & 25.19\% \\
\textbf{Supporting A}  &  13.45\%           &  8.28\%   &   \textbf{29.34\%}  &  7.59\%  & 12.59\%  \\
\textbf{Supporting B}  &  10.02\%           &  \textbf{32.41\%}  &   14.67\%  &  \textbf{45.16\%} & 17.04\%  \\
\textbf{Supporting C}  &  14.80\%           &  20.69\%  &   13.33\%  &  19.66\% & \textbf{34.44\%}  \\ \midrule
\textbf{Total}         &  100\%             &  100\%  &   100\%  &  100\% & 100\%  \\ \bottomrule
\end{tabular}%
}
\caption{Accuracy of perceived speaking style in speech samples generated from different VAE z-vector centroids.}
\label{table:controllability}
\end{table}

In this experiment, we investigate the controllability of speaking style via the reference encoder.
Figure \ref{fig:vae_space} illustrates the VAE space for the systems augmented with conversational data from 4 speakers and 8 speakers, described in section \ref{subsec:supporting_speakers}.
We achieve a reasonable separation of speaking styles, represented by clusters corresponding to individual speakers.
Taking the system augmented with data from 4 speakers, we synthesize a set of samples conditioned on each of the VAE centroids, computed over the voice-converted training data from each speaker.
We ask listeners to rate samples in terms of speaking style similarity against a natural reference produced by the target or supporting speakers.
For each reference speaker, 30 listeners rated a total of 50 utterances in batches of 10 MUSHRA screens.
For this evaluation, we asked listeners to ignore the identity of the voice and to focus solely on the similarity of the speaking style.
To simplify our analysis, we took the highest rated sample for each submission to be the perceived speaking style.
We then computed the accuracy for each centroid with respect to the reference speaker.

\textbf{Results} are summarized in Table \ref{table:controllability} and indicate that listeners tend to perceive the correct speaking style.
However, overall accuracy scores are still lower than expected and show some confusion across reference speakers.
We hypothesize that this might be due to an inherent speaking style similarity across supporting speakers.
Further work should validate these observations, evaluating systems with styles that are more perceptually different.

\subsection{Amount of supporting data}
\label{subsec:amount_data}

We investigate the amount of supporting data required for cross-speaker style transfer.
As before, we train a TTS system on 10 hours of data in a neutral speaking style from the target speaker, termed \emph{Neutral}.
We also train a system on 8 hours of conversational data from the \emph{Source speaker}.
The cross-speaker style transfer systems are augmented with conversational data from 4 supporting speakers.
We vary the total amount of available supporting conversational data, considering 8h, 6h, 3h, and 1h of data.
The amount of data is distributed equally across the 4 conversational speakers.
Therefore, we use 2 hours of data per speaker when considering a total of 8 hours of supporting data;
and we take 15 minutes of data per speaker for the system augmented with 1 hour of data.

We follow the evaluation methodology described in section \ref{subsec:supporting_speakers}.
\textbf{Results} are summarized in Table \ref{table:exp_amount_data}.
We conduct two-sided t-tests on the MUSHRA scores with a Holm-Bonferroni correction on the three evaluations.
For all listening tests, we observe no statistically significant differences between the augmented systems.
For these systems, in terms of speaker similarity, we reduce the gap between source and target speaker systems by 93.3\%.
As before, augmented systems outperform the neutral system for style similarity.
These results suggest that the proposed approach is effective with a reduced amount of supporting data.

\begin{table}[t]
\centering
\resizebox{1.0\columnwidth}{!}{%
\begin{tabular}{lccc}
\toprule
\textbf{System}     & \textbf{Naturalness}  & \textbf{Speaker Sim.} & \textbf{Style Sim.}           \\ \midrule
Recordings          &  64.69 $\pm$ 1.48     &    -                  & -                   \\
Neutral             &  56.71 $\pm$ 1.42     &    72.32 $\pm$ 1.48   & 40.21 $\pm$ 1.69    \\
Data (1 hour)       &  57.82 $\pm$ 1.35     &    69.25 $\pm$ 1.45   & 55.41 $\pm$ 1.57    \\
Data (3 hours)      &  57.58 $\pm$ 1.34     &    69.84 $\pm$ 1.46   & 56.68 $\pm$ 1.57    \\
Data (6 hours)      &  58.19 $\pm$ 1.35     &    68.84 $\pm$ 1.48   & 56.03 $\pm$ 1.58    \\
Data (8 hours)      &  56.55 $\pm$ 1.37     &    69.51 $\pm$ 1.48   & 55.49 $\pm$ 1.59    \\
Source Speaker      &  -                    &    28.33 $\pm$ 1.65   & 73.69 $\pm$ 1.40     \\ \bottomrule
\end{tabular}%
}
\caption{MUSHRA evaluation in terms of Naturalness, Speaker Similarity, and Speaking Style Similarity. Results indicate mean score with 95\% confidence interval. \enquote{Data} systems denote TTS systems augmented with $n$ hours of conversational data distributed equally across 4 supporting speakers. \enquote{Source Speaker} indicates a conversational TTS system trained on 8 hours from the source speaker.}
\label{table:exp_amount_data}
\end{table}

\subsection{Does it scale?}
\label{subsec:scalability}

We validate our proposed approach on 14 different speakers, distributed across 7 dialects.
We restrict the training data of each target speaker to a maximum of 10 hours of speech in a neutral reading style.
For supporting data in a conversational speaking style, we consider 3 supporting speakers, each contributing with 1 hour of conversational data.
\emph{Augmented} systems are trained following the pipeline described in section \ref{sec:methodology}.
\emph{Neutral} systems are trained using only the 10 hours of single-speaker neutral data.
We evaluate systems in terms of speaker and style similarity.
For speaker similarity, we follow the same paradigm as described before, although for simplicity we replace the synthetic source speaker sample with a vocoded sample.
For style similarity, the reference is a sample from the source speaker in a conversational speaking style, matching the text of the synthetic utterances.
We consider the neutral and augmented systems, and include as topline the corresponding voice converted sample.
For each evaluation, 100 native speakers rated a set of 150 utterances, with each participant being assigned no more than 15 MUSHRA screens.

\textbf{Results} are presented in Table \ref{table:exp_scalability}.
In terms of speaker similarity, Augmented systems, on average, bridge the gap by 91.42\% relative to the lower-anchor Source Speaker and the upper-anchor Neutral system (\enquote{Rel} column for speaker similarity in Table \ref{table:exp_scalability}).
We observe that only four systems bridge the gap by less than 90\%.
For style similarity, we consider the Neutral system to be the lower-anchor and the Voice Converted samples to be the upper-anchor.
In this case, Augmented systems bridge the gap, on average, by 57.6\% (\enquote{Rel} column for style similarity in Table \ref{table:exp_scalability}).
If we instead consider the Reference sample as our upper-anchor (assumed to be 100), then the Augmented systems bridge the gap, on average, by 18.5\% over the Neutral samples.
We additionally note that the difference between Augmented and Neutral systems is statistically significant at the level of $p<.01$ for all systems, except the German Male voice.
Results indicate that our proposed approach is generally successful when transferring speaking style while preserving speaker identity.
Typically, female target speakers perform better than male target speakers, which is likely due to a difference in speaking style between supporting and target speakers.
Nonetheless, our results are extremely positive, in particular considering that we use only 3 hours of expressive data, with only 1 hour from the source speaker.

\begin{table}[t]
\centering
\resizebox{1.0\columnwidth}{!}{%
\begin{tabular}{lcllll|llll}
\toprule
\multirow{2}{*}{\textbf{Lang}}    & \multirow{2}{*}{\textbf{Gend}} &  \multicolumn{4}{c}{\textbf{Speaker Sim}} & \multicolumn{4}{c}{\textbf{Style Sim}} \\ \cmidrule(l){3-10}
              &                  & Src     & Neut    & Aug    & \emph{Rel}  & VC      & Neut    & Aug     & \emph{Rel}   \\ \midrule
\multirow{2}{*}{pt-PT}  &  F     & 24.65   & 74.46   & 59.17  & 69.30\%     & 68.89   & 46.85   & 65.22   & 83.35\%       \\
                        &  M     & 15.52   & 78.94   & 74.10  & 92.37\%     & 65.75   & 49.57   & 62.65  & 80.84\%       \\ \cmidrule(l){2-10}

\multirow{2}{*}{pt-BR}  &  F     & 26.17   & 79.32  & 67.81   & 78.34\%     & 69.58   & 43.25   & 64.55   & 80.90\%       \\
                        &  M     & 19.55   & 90.07  & 77.25   & 82.00\%     & 70.12   & 56.41   & 63.98   & 55.22\%       \\ \cmidrule(l){2-10}

\multirow{2}{*}{es-ES}  &  F     & 35.67   & 76.53  & 71.94   & 88.77\%     & 71.14   & 53.20   & 65.89   & 70.74\%       \\
                        &  M     & 21.91   & 79.29  & 76.39   & 94.95\%     & 73.41   & 57.75   & 67.89   & 64.75\%       \\ \cmidrule(l){2-10}

\multirow{2}{*}{es-MX}  &  F     & 47.66   & 65.67  & 65.58   & 99.50\%     & 70.98   & 59.61   & 67.91   & 73.00\%       \\
                        &  F     & 49.77   & 70.40  & 70.85   & 102.10\%     & 68.95   & 63.07   & 65.61  & 43.20\%       \\ \cmidrule(l){2-10}

\multirow{2}{*}{de-DE}  &  F     & 26.40   & 77.30  & 73.11   & 91.77\%     & 69.02   & 56.64   & 62.12   & 44.26\%       \\
                        &  M     & 27.48   & 74.14  & 72.33   & 96.10\%      & 64.68   & 55.73   & 57.20  & 16.42\%       \\ \cmidrule(l){2-10}

\multirow{2}{*}{it-IT}  &  F     & 25.67   & 69.42  & 67.25   & 95.00\%     & 63.23   & 48.38   & 55.36   & 47.00\%       \\
                        &  M     & 17.70    & 81.29 & 80.39   & 98.60\%     & 71.22   & 54.19   & 59.20   & 29.42\%       \\ \cmidrule(l){2-10}

\multirow{2}{*}{fr-CA}  &  F     & 36.87   & 69.47  & 68.12   & 95.85\%     & 65.87   & 55.57   & 61.26   & 55.24\%       \\
                        &  F     & 38.10   & 68.75  & 67.31   & 95.30\%     & 64.22   & 55.64   & 60.97   & 62.12\%       \\

\bottomrule
\end{tabular}%
}
\caption{Cross-speaker style transfer for 14 speakers (9 female, 5 male). Results indicate mean MUSHRA score for speaker and style similarity. \enquote{Rel} indicates the relative position of the Augmented system (\enquote{Aug}) to the remaining two competing systems: Neutral (\enquote{Neut}), source speaker (\enquote{Src}), or voice converted sample (\enquote{VC}).}
\label{table:exp_scalability}
\end{table}

\section{Conclusion and future work}

We addressed the problem of cross-speaker style transfer for TTS using data augmentation.
Our approach uses a voice conversion model to generate high-quality data from a set of supporting expressive speakers.
The voice converted samples are pooled with natural data from the target speaker to train a single-speaker multi-style TTS system.
Results indicate that our proposed method works well when using a single or multiple supporting speakers, achieving good results with as little as 1 hour of expressive supporting data.
Future work will focus on the development of the respective voice conversion and text-to-speech architectures.
With respect to the TTS model architecture, we will investigate better disentanglement and stronger controllability of speaking style from the supporting conversational data.
Additionally, we aim to scale the proposed methodology to more expressive speaking styles, such as emotions.
Overall experimental results show that our proposed approach is efficient and scalable to multiple languages when transfer speaking style.